\documentclass[sigconf]{acmart}
\AtBeginDocument{%
  \providecommand\BibTeX{{%
    \normalfont B\kern-0.5em{\scshape i\kern-0.25em b}\kern-0.8em\TeX}}}

\usepackage{siunitx}
\usepackage{enumitem}
\usepackage{xurl}
\usepackage[htt]{hyphenat}
\usepackage{graphicx}

\usepackage[acronym,shortcuts]{glossaries}
\newacronym{ARMD}{ARMD}{age-related macular degeneration}
\newacronym{FOV}{FOV}{field of view}
\newacronym{HCI}{HCI}{human-computer interaction}
\newacronym{HMD}{HMD}{head-mounted display}
\newacronym{NFB}{NFB}{nerve fiber bundle}
\newacronym{RGC}{RGC}{retinal ganglion cell}
\newacronym{RP}{RP}{retinitis pigmentosa}
\newacronym{SPV}{SPV}{simulated prosthetic vision}
\newacronym{VPU}{VPU}{video processing unit}
\newacronym{VR}{VR}{virtual reality}

\setcopyright{acmcopyright}
\copyrightyear{2022}
\acmYear{2022}
\acmDOI{10.1145/3562939.3565620}

\copyrightyear{2022}
\acmYear{2022}
\setcopyright{rightsretained}
\acmConference[VRST '22]{28th ACM Symposium on Virtual Reality Software and Technology}{November 29-December 1, 2022}{Tsukuba, Japan}
\acmBooktitle{28th ACM Symposium on Virtual Reality Software and Technology (VRST '22), November 29-December 1, 2022, Tsukuba, Japan}
\acmDOI{10.1145/3562939.3565620}
\acmISBN{978-1-4503-9889-3/22/11}

\begin{document}

\title[Depth Cues and Semantic Edges for Indoor Mobility Using Simulated Prosthetic Vision]{The Relative Importance of Depth Cues and Semantic Edges for Indoor Mobility Using Simulated Prosthetic Vision in Immersive Virtual Reality}


\author{Alex Rasla}
\affiliation{%
  \institution{University of California, Santa Barbara}
  \city{Santa Barbara}
  \country{CA, USA}}
\email{alexrasla@ucsb.edu}

\author{Michael Beyeler}
\affiliation{%
  \institution{University of California, Santa Barbara}
  \city{Santa Barbara}
  \country{CA, USA}
}
\email{mbeyeler@ucsb.edu}

\renewcommand{\shortauthors}{Rasla \& Beyeler}

\begin{abstract}
Visual neuroprostheses (\emph{bionic eyes}) have the potential to treat degenerative eye diseases that often result in low vision or complete blindness.
These devices rely on an external camera to capture the visual scene, which is then translated frame-by-frame into an electrical stimulation pattern that is sent to the implant in the eye. 
To highlight more meaningful information in the scene, recent studies have tested the effectiveness of deep-learning based computer vision techniques, such as depth estimation to highlight nearby obstacles (\emph{DepthOnly} mode) and semantic edge detection to outline important objects in the scene (\emph{EdgesOnly} mode).
However, nobody has yet attempted to combine the two, either by presenting them together (\textit{EdgesAndDepth}) or by giving the user the ability to flexibly switch between them (\textit{EdgesOrDepth}).
Here, we used a neurobiologically inspired model of simulated prosthetic vision (SPV) in an immersive virtual reality (VR) environment to test the relative importance of semantic edges and relative depth cues to support the ability to avoid obstacles and identify objects.
We found that participants were significantly better at avoiding obstacles using depth-based cues as opposed to relying on edge information alone, and that roughly half the participants preferred the flexibility to switch between modes (\emph{EdgesOrDepth}).
This study highlights the relative importance of depth cues for SPV mobility and is an important first step towards a visual neuroprosthesis that uses computer vision to improve a user's scene understanding.
\end{abstract}



\begin{teaserfigure}
\centering
  \includegraphics[width=.86\textwidth]{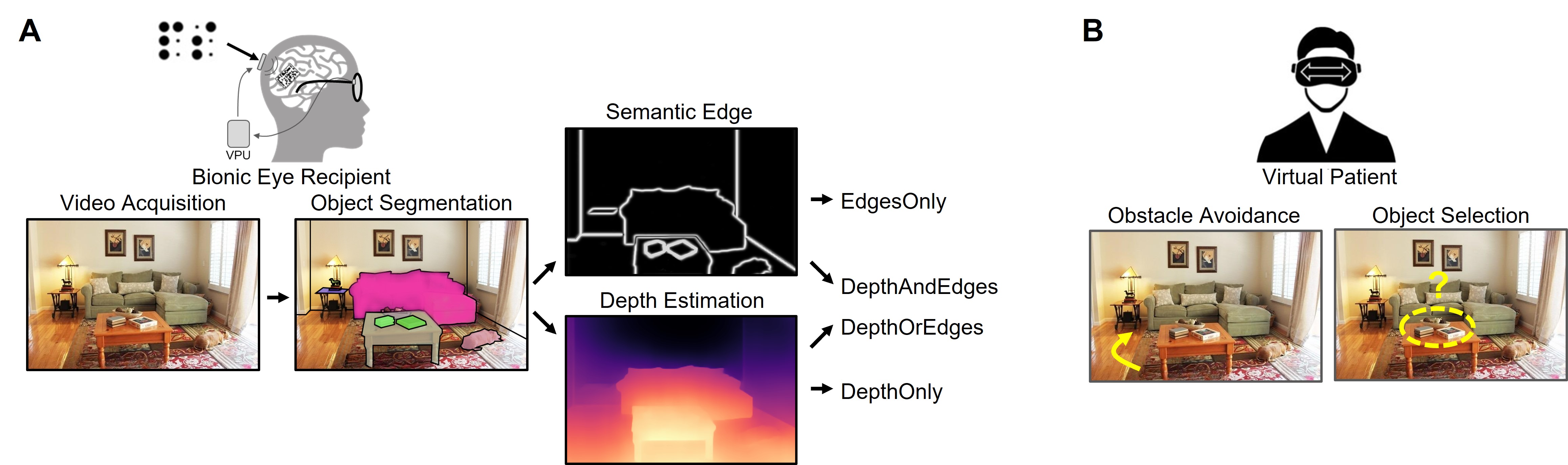}
  \Description{Scene simplification for bionic vision. Panel A shows a flow diagram going from image acquisition to the four different augmentation modes. 
  The original image shows a living room, consisting of a couch and a lamp on a sidetable; in front of them is a coffee table with several objects on it. First, objects are segmented in the original image. Then the object boundaries are highlighted as semantic edges, and depth is extracted. Panel B shows the schematic of a virtual patient wearing a head-mounted display. An arrow outlines a path around the obstacle in the image (obstacle avoidance) and several objects on the coffee table (object selection).}
  \caption{Scene simplification for bionic vision.
  A) Visual neuroprostheses (\emph{bionic eyes}) electrically stimulate neurons in the visual system to restore a rudimentary form of vision to people living with incurable blindness (\emph{inset}).
  To create meaningful artificial vision, the visual scene is simplified by extracting semantic edges and estimating relative depth before it is displayed, here illustrated on an indoor scene from the MS-COCO database.
  Semantic edges and depth cues may be visualized either independently (\emph{EdgesOnly} and \emph{DepthOnly} mode) or together (\emph{EdgesAndDepth}).
  Alternatively, users may prefer the ability to flexibly switch between edges and depth cues (\emph{EdgesOrDepth}).
  B) As a proof of concept, we used a neurobiologically inspired computational model to generate realistic predictions of simulated prosthetic vision, and asked sighted subjects (i.e., \emph{virtual patients}) to avoid obstacles and select objects in an immersive virtual reality environment.}
  \label{fig:overview}
\end{teaserfigure}

\maketitle

\section{Introduction}

By the year 2050, roughly 114.6 million people will be living with incurable blindness \cite{bourne_magnitude_2017}.
Although some individuals can be treated with surgery or medication, there are no effective treatments for many people blinded by severe degeneration or damage to the retina, the optic nerve, or cortex. In such cases, an electronic visual prosthesis (\emph{bionic eye}) may be the only option \cite{fernandez_development_2018} (Fig.~\ref{fig:overview}A)
Analogous to cochlear implants, these devices electrically stimulate neurons in the early visual system to elicit neuronal responses that the brain interprets as visual percepts (\emph{phosphenes}).

Current devices generally provide users with an improved ability to localize high-contrast objects and perform basic orientation \& mobility tasks \cite{ayton_update_2020}, but are not yet able to match the acuity of natural vision.
Most current prostheses provide a very limited \ac{FOV}; for example, the artificial vision generated by Argus II~\cite{luo_argusr_2016}, the most widely adopted retinal implant thus far, is restricted to roughly $10 \times 20$ degrees of visual angle. This forces users to scan the environment with strategic head movements while attempting to piece together the information \cite{erickson-davis_what_2021}.
In addition, the limited number of electrodes (60 in Argus II) severely limits the number of independent phosphenes that the device can generate \cite{beyeler_learning_2017}.

Consequently, researchers have suggested ways to simplify the visual scene before it is displayed using image processing and computer vision.
One popular approach is to estimate relative depth in the scene \cite{mccarthy_mobility_2014,han_deep_2021} and then make phosphenes appear brighter the closer they are to the observer, in order to highlight nearby obstacles.
Here we refer to this method of substituting depth for intensity as \emph{DepthOnly} mode (Fig.~\ref{fig:overview}A).
A more recent line of research suggests to extract semantic and structural edges instead \cite{sanchez-garcia_indoor_2019,sanchez-garcia_semantic_2020} (\emph{EdgesOnly} mode), in order to give the user a sense of where important objects are in the scene.

Since these two modes provide seemingly complementary sources of information, it is natural to ask how to best combine them.
A straightforward approach would be to visualize both edge and depth cues at the same time (\emph{EdgesAndDepth}).
However, we hypothesized that users might instead prefer the ability to flexibly switch between the two (\emph{EdgesOrDepth}).
To assess the relative importance of depth and edge information, a side-by-side comparison is needed.

Due to the unique requirements of working with bionic eye recipients (e.g., constant assistance, increased setup time, travel cost), experimentation with different encoding methods remains challenging and expensive.
Instead, embedding models of \ac{SPV} in immersive \ac{VR} allows sighted subjects to act as \emph{virtual patients} by ``seeing'' through the eyes of the patient, taking into account their head and eye movements as they explore an immersive virtual environment \cite{kasowski_immersive_2022,wang_cross-task_2018,thorn_virtual_2020,sanchez-garcia_influence_2020}.
This can speed up the development process by allowing researchers to test theoretical predictions in high-throughput experiments, the best of which can be validated and improved upon in an iterative process with the bionic eye recipient in the loop \cite{kasowski_towards_2021}.

To this end, we make the following contributions:
\begin{enumerate}[topsep=0pt,itemsep=-1ex,partopsep=0pt,parsep=1ex,leftmargin=14pt,label=\roman*.]
    \item We embed a psychophysically validated \ac{SPV} model \cite{beyeler_model_2019} in \ac{VR} to allow sighted participants to act as virtual bionic eye patients in an immersive virtual environment.
    \item We explore the relative importance of different scene simplification strategies based on depth estimation \cite{han_deep_2021} and semantic edge detection \cite{sanchez-garcia_semantic_2020} as a preprocessing strategy for bionic vision (Fig.~\ref{fig:overview}A).
    \item We systematically evaluate the ability of these strategies to support obstacle avoidance and object identification (Fig.~\ref{fig:overview}B) with a user study in immersive \ac{VR}.
\end{enumerate}

In sum, this is the first study to compare the relative importance of depth cues and semantic edges for bionic vision and an essential first towards a bionic eye that uses computer vision to improve a user's scene understanding.

\section{Background}
\label{sec:background}

Retinal implants are currently the only FDA-approved technology to treat blinding degenerative diseases such as \ac{RP} and \ac{ARMD}.
Most current devices acquire visual input via an external camera and perform edge extraction or contrast enhancement via an external \ac{VPU}, before sending the signal through wireless coils to a microstimulator implanted in the eye or the brain (see Fig.~\ref{fig:overview}A).
This device receives the information, decodes it, and stimulates the visual system with electrical current.
Two devices are already approved for commercial use: Argus II (60 electrodes arranged in a $6 \times 10$ grid, Second Sight Medical Products, Inc., \cite{luo_argusr_2016}) and Alpha-IMS (1500 electrodes, Retina Implant AG, \cite{stingl_artificial_2013}).
In addition, PRIMA (378 electrodes, Pixium Vision, 
has started clinical trials, with others to follow shortly \cite{ayton_first--human_2014,ferlauto_design_2018}.

A common misconception is that each electrode in the grid can be thought of as a ``pixel'' in an image \cite{dagnelie_real_2007,chen_simulating_2009,lui_transformative_2011,perez-yus_depth_2017,sanchez-garcia_indoor_2019}, and most retinal implants linearly translate the grayscale value of a pixel in each video frame to a current amplitude of the corresponding electrode in the array \cite{luo_argusr_2016}.
This is known as the scoreboard model, which implies that creating a complex visual scene can be accomplished simply by using the right combination of pixels, analogous to the images projected on the light bulb arrays of some sports stadium scoreboards \cite{dobelle_artificial_2000}.
On the contrary, recent work suggests that phosphenes vary in shape and size, and differ substantially across subjects and electrodes \cite{beyeler_model_2019,erickson-davis_what_2021}. 

Despite their potential to restore vision to people living with incurable blindness, the number of bionic eye users in the world is still relatively small ($\sim 500$ retinal prostheses implanted to date).
To investigate functional recovery and experiment with different implant designs, researchers have therefore been developing \ac{VR} prototypes that rely on \ac{SPV}.
The classical method relies on sighted subjects wearing a \ac{VR} headset, who are then deprived of natural viewing and only perceive phosphenes displayed in a \ac{HMD}. This viewing mode has been termed \emph{transformative reality} \cite{lui_transformative_2011} (as opposed to \emph{altered reality} typically used to describe low vision simulations \cite{bao_augmented_2019}), which allows sighted users to ``see'' through the eyes of the bionic eye recipient, taking into account their head and/or eye movements as they explore a virtual environment \cite{kasowski_towards_2021}.

However, because most \ac{SPV} studies rely on the scoreboard model \cite{dagnelie_real_2007,chen_simulating_2009,lui_transformative_2011,perez-yus_depth_2017,sanchez-garcia_indoor_2019}, it is unclear how their findings would translate to real bionic eye recipients.
Only a handful of studies have incorporated a great amount of neurophysiological detail into their setup~\cite{josh_psychophysics_2013,vurro_simulation_2014,wang_cross-task_2018,thorn_virtual_2020,kasowski_immersive_2022}, only three of which relied on an established and psychophysically validated model of \ac{SPV} \cite{wang_cross-task_2018,thorn_virtual_2020,kasowski_immersive_2022}.
In addition, being able to move around as one would in real life has shown to significantly increase the level of immersion a user experiences \cite{pasch_immersion_2009}.
However, the level of immersion offered by most \ac{SPV} studies is relatively low, as stimuli are often presented on a screen \cite{ying_recognition_2018, wang_cross-task_2018}.
In contrast, most current prostheses provide a very limited \ac{FOV} (e.g., Argus II: $10 \times 20$ degrees of visual angle), which requires users to scan the environment with strategic head movements while trying to piece together the information \cite{erickson-davis_what_2021}.
Furthermore, Argus II does not take into account the eye movements of the user when updating the visual scene, which can be disorienting for the user.
Ignoring these \ac{HCI} aspects of bionic vision can result in unrealistic predictions of prosthetic performance, sometimes even exceeding theoretical limits for visual acuity (as pointed out by \cite{caspi_assessing_2015}).

\section{Related Work}
\label{sec:related_work}

Most retinal implants are equipped with an external \ac{VPU} that is capable of applying simple image processing techniques to the video feed in real time.
In the near future, these techniques may include deep learning--based algorithms aimed at improving a patient's scene understanding \cite{beyeler_towards_2022}.
Based on this premise, researchers have developed various image optimization strategies, and assessed their performance by having sighted observers (i.e., \emph{virtual patients}) conduct daily visual tasks under \ac{SPV} \cite{boyle_region--interest_2008,dagnelie_real_2007,al-atabany_improved_2010,li_image_2018,mccarthy_mobility_2014,vergnieux_simplification_2017}.
These simulations allow a wide range of computer vision systems to be developed and tested without requiring implanted devices.

Current retinal prostheses are implanted in only one eye, and thus are unable to convey binocular depth cues.
Previous work has therefore explored the possibility of obtaining depth information through additional peripherals, such as an RGB-D sensor, and studied behavioral performance of virtual patients typically navigating an obstacle course under \ac{SPV}.
For example, Ref.~\cite{perez-yus_depth_2017} used depth cues to generate a simplified representation of the ground to indicate the free space within which virtual patients could safely walk around, whereas Ref.~\cite{han_deep_2021} used deep neural networks to estimate per-pixel relative depth and then substituted depth for intensity.
Depth cues were also shown to help avoid nearby obstacles that are notoriously hard to detect with other computer vision algorithms, such as branches hanging from a tree \cite{lieby_substituting_2011}.
Ref.~\cite{mccarthy_mobility_2014} used depth to increase the contrast of object boundaries and showed that this method reduced the number of collisions with ground obstacles.
In addition, retinal prosthesis patients were shown to benefit from distance information provided by a thermal sensor when trying to avoid nearby obstacles and people \cite{sadeghi_thermal_2019}.

Recently, with the development of deep learning in computer vision, semantic segmentation algorithms have become unprecedentedly effective. This can be used as another method to reduce visual clutter in prosthetic vision, be it in  outdoor scenes \cite{horne_semantic_2016} or indoor scenes \cite{sanchez-garcia_indoor_2019}.
The latter study combined semantic and structural image segmentation to build a schematic representation of indoor environments, which was then shown to improve object and room identification in a \ac{SPV} task \cite{sanchez-garcia_indoor_2019}. 
 Semantic segmentation has been applied to simplify both the outdoor scenes \cite{horne_semantic_2016} and the indoor scenes \cite{sanchez-garcia_indoor_2019} for retinal prostheses.

\begin{figure*}[t!]
    \centering
    \includegraphics[width=\linewidth]{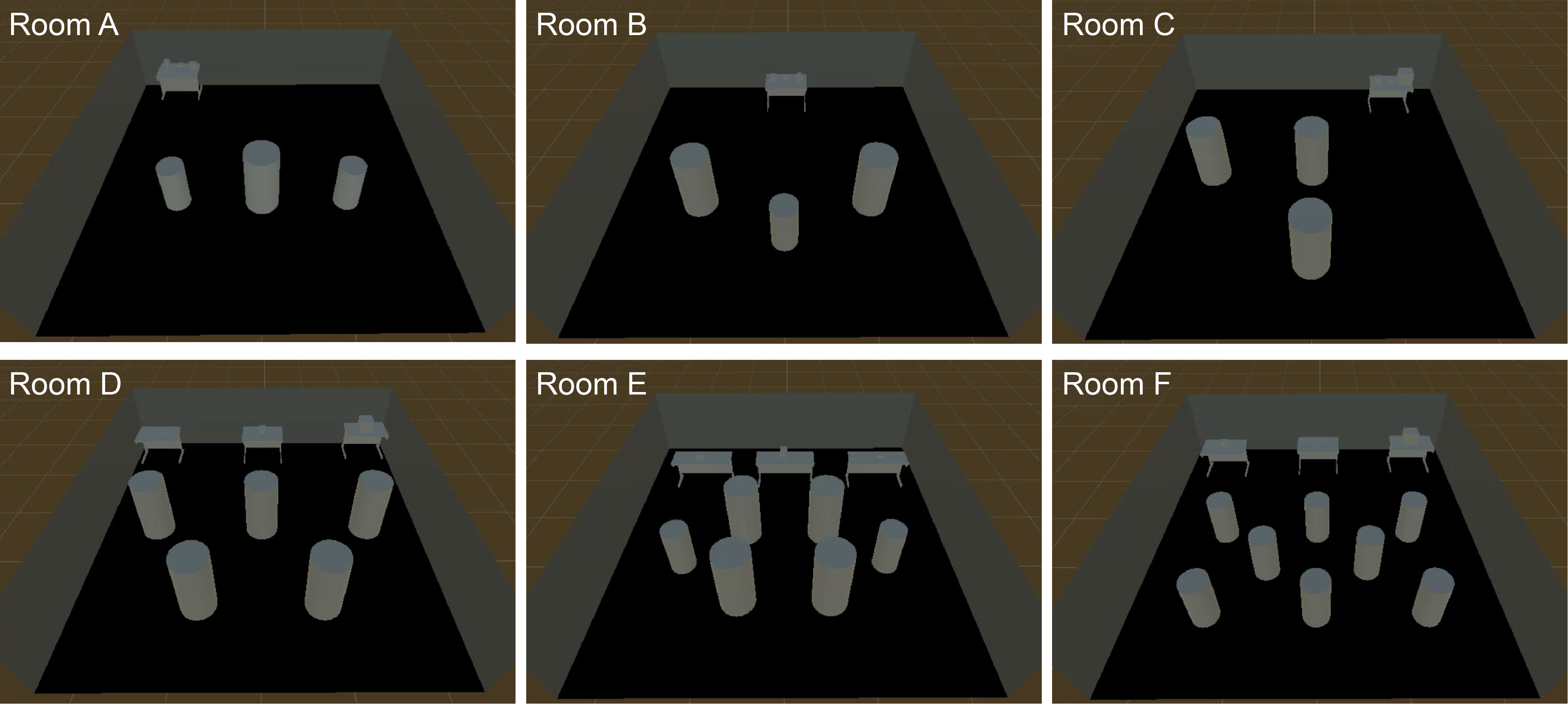}
    \Description{Six different room layouts shown from above. Bottom half of each room contains either three cylinders (Rooms A - C) arranged differently, five cylinders (Room D), six cylinder (Rooms E), or eight cylinders (Room F) arranged in hexagonal grid.  At the end of the room, there was either one table with three objects on it, or three tables with one object on each.}
    \caption{Room layouts. Participants started in the center along the bottom wall.
    Participants were instructed to walk towards the other end of the room while avoiding obstacles.
    At the end of the room, there was either one table with three objects on it, or three tables with one object on each.
    Participants had to identify the medium-sized cube located one on of the tables.}
    \label{fig:rooms}
\end{figure*}

However, since the above algorithms were developed in isolation and tested on different behavioral tasks, a side-by-side comparison of their ability to aid scene understanding is still lacking.
A notable exception is Ref.~\cite{han_deep_2021}, which compared a number of scene simplification strategies that rely on depth extraction, semantic segmentation, and visual saliency on a single task.
However, their study was limited to sighted participants viewing \ac{SPV} videos on a screen, and thus did not account for the \ac{FOV} restrictions that are common with current implants. 

To address these challenges, we used a neurobiologically inspired computational model of bionic vision \cite{beyeler_model_2019} to generate realistic predictions of \ac{SPV}, and combined it with scene simplification strategies based on depth estimation \cite{han_deep_2021} and semantic edge detection \cite{sanchez-garcia_semantic_2020}.
To allow for a fair comparison between algorithms, we asked virtual patients to avoid obstacles and identify objects in a number of immersive virtual environments.

\section{Methods}

\subsection{Virtual Patients}

To simulate a bionic eye patient, we developed \ac{SPV} simulations in Unity that were streamed in real time to a wireless head-mounted \ac{VR} headset (HTC VIVE Pro Eye with wireless adapter, HTC Corporation).
We followed the procedure outlined in Fig.~\ref{fig:overview}A to simulate different scene simplification strategies and tested them on an obstacle avoidance and object selection task (Fig.~\ref{fig:overview}B).
All our simulations were run on an Intel i9-9900k processor (C\# code) and an NVIDIA RTX 2070 Super GPU with 16GB of DDR4 memory (shader code).
The entire \ac{SPV} workflow was thus as follows:
\begin{enumerate}[topsep=0em,itemsep=-1ex,partopsep=0em,parsep=1ex,leftmargin=6ex,label=\roman*.]
    \item Image acquisition: Utilize Unity's virtual camera to acquire the central $60 \times 60$ degrees of visual angle of the display's monocular \ac{FOV} at roughly 90 frames per second and downscale to a target texture of $120 \times 120$ pixels.
    \item Scene simplification: Extract semantic edges using the \texttt{QuickOutline} asset and/or depth cues using \texttt{UnityObjectToClipPos} to simulate different scene simplification strategies (explained in Section~\ref{sec:modes}).
    \item Electrode activation: Determine electrode activation based on the visual input as well as the placement of the simulated retinal implant. In the current study, a $3 \times 3$ Gaussian blur was applied to the preprocessed image to average the grayscale values around each electrode's location in the visual field. This gray level was then interpreted as a current amplitude delivered to a particular electrode in the array.
    \item Phosphene model: Use Unity shaders to convert electrode activation to a visual scene in real time (see Section~\ref{sec:spv}).
    \item Phosphene rendering: Render the elicited phosphenes in the \ac{HMD} of the \ac{VR} system.
\end{enumerate}

\subsection{Participants}

We recruited $18$ participants with normal or corrected-to-normal vision from the research participant pool of the Department of Psychological \& Brain Sciences at University of California, Santa Barbara (UCSB) to act as virtual patients in our experiment.
Participants ranged from 18 to 20 years in age ($\mathit{M}=18.8$, $\mathit{SD}=.73$ years), with 6 participants identifying as male and 12 identifying as female.
Of these participants, $4$ had never used \ac{VR} before, $12$ had used VR $1-5$ times before, $1$ had used it $10-20$ times, and $1$ had used it $20+$ times.
Potential participants were excluded if they reported that they were prone to cybersickness.
The study was approved by UCSB's Institutional Review Board.

\subsection{Rooms}
\label{sec:rooms}

Participants were asked to navigate six different virtual rooms that were filled with 3-7 obstacles to avoid and a target object to select among distractor objects (Fig.~\ref{fig:rooms}).
Participants always started along the bottom wall and were instructed to walk towards the other end of the room while avoiding obstacles.
Upon collision with an object, a ``thud'' sound was played through the VIVE headphones.

\begin{figure}[ht]
    \centering
    \includegraphics[width=\linewidth]{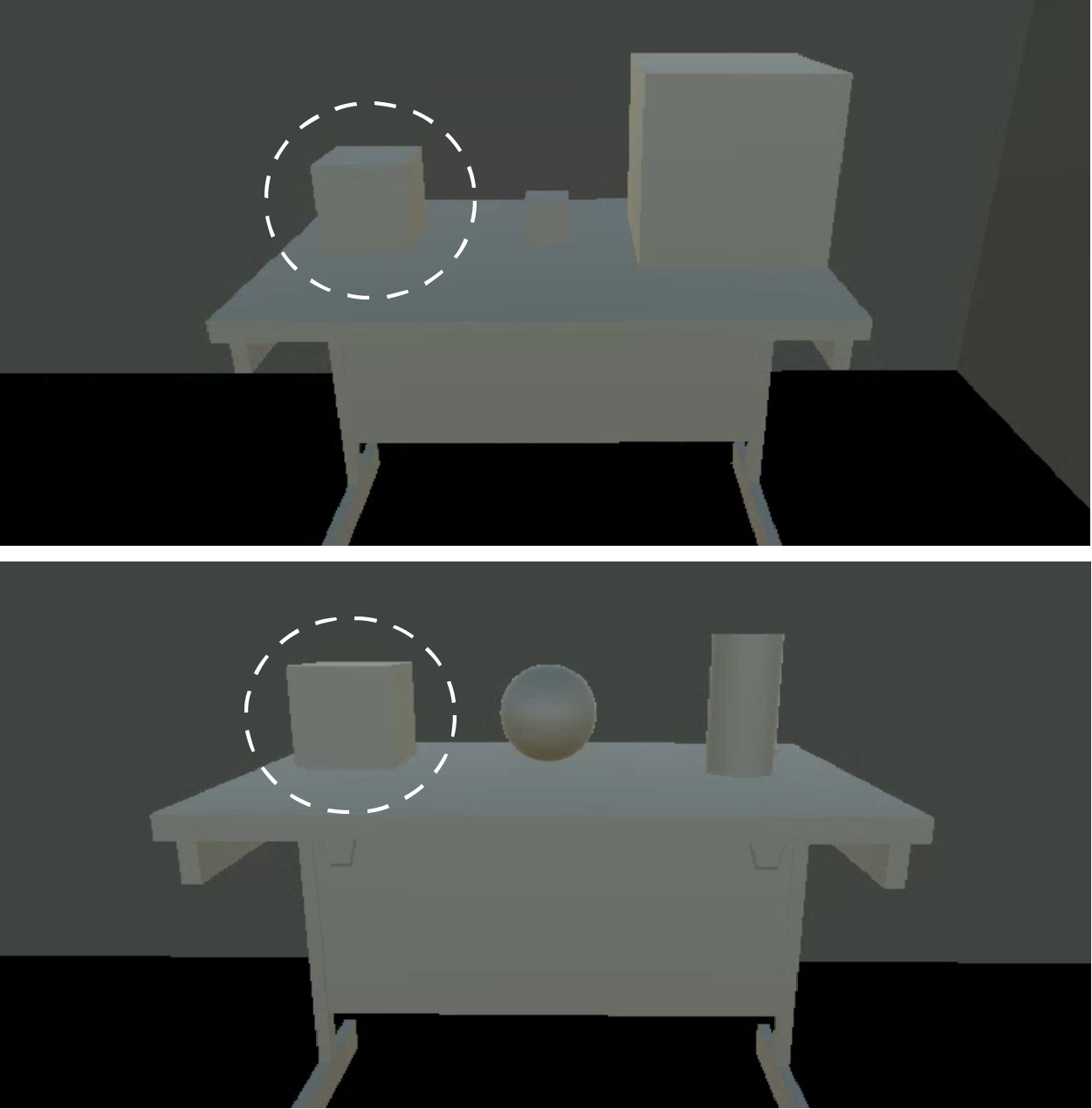}
    \Description{Two example views of the table with different objects on it. Top panel shows three different boxes, bottom panel shows a box, a sphere, and a cylinder.}
    \caption{Example views of the table with different objects on it. The correct object to identify was always the medium-sized cube (indicated with a white dashed circle) among two distractor objects.
    The arrangement of the objects was pseudo-randomized on each trial.
    Users had to confirm their selection by pressing a button on the VIVE controllers.}
    \label{fig:rooms-table}
\end{figure}

Once they had passed all obstacles, a ``chime'' sound was played through the VIVE headphones to indicate the end of the obstacle avoidance portion of the task.
Participants then had to navigate to one of three tables and identify a medium-sized cube located on it (Fig.~\ref{fig:rooms-table}).
Each room had either one table with three objects on it, or three tables with one object each.
While the arrangements of objects was pseudo-randomized on each trial (consisting either of a sphere, a cylinder, and a  medium cube; or a small cube, a medium cube, and a large cube), participants always had to select the medium cube.

\subsection{Simulated Prosthetic Vision}
\label{sec:spv}

To generate realistic simulations of phosphene appearance, we adapted code from the VR-SPV open-source toolbox \cite{kasowski_immersive_2022}.
This toolbox provides a C\#/shader implementation of the pyschophysically validated axon map model \cite{beyeler_model_2019} that describes phosphenes using two shape parameters, $\rho$ (specifying phosphene size) and $\lambda$ (describing phosphene elongation).
We chose $\rho=300$ and $\lambda=550$ for our simulations, which is roughly in the middle of the range of values reported by Argus II users \cite{beyeler_model_2019}.
To support the real-time simulation of a $20 \times 15$ electrode array implanted in the retina, we pre-calculated an initial mapping of each electrode's effects on the scene before starting the experiment.
 
\subsection{Modes}
\label{sec:modes}

\begin{figure}[ht]
    \centering
    \includegraphics[width=\linewidth]{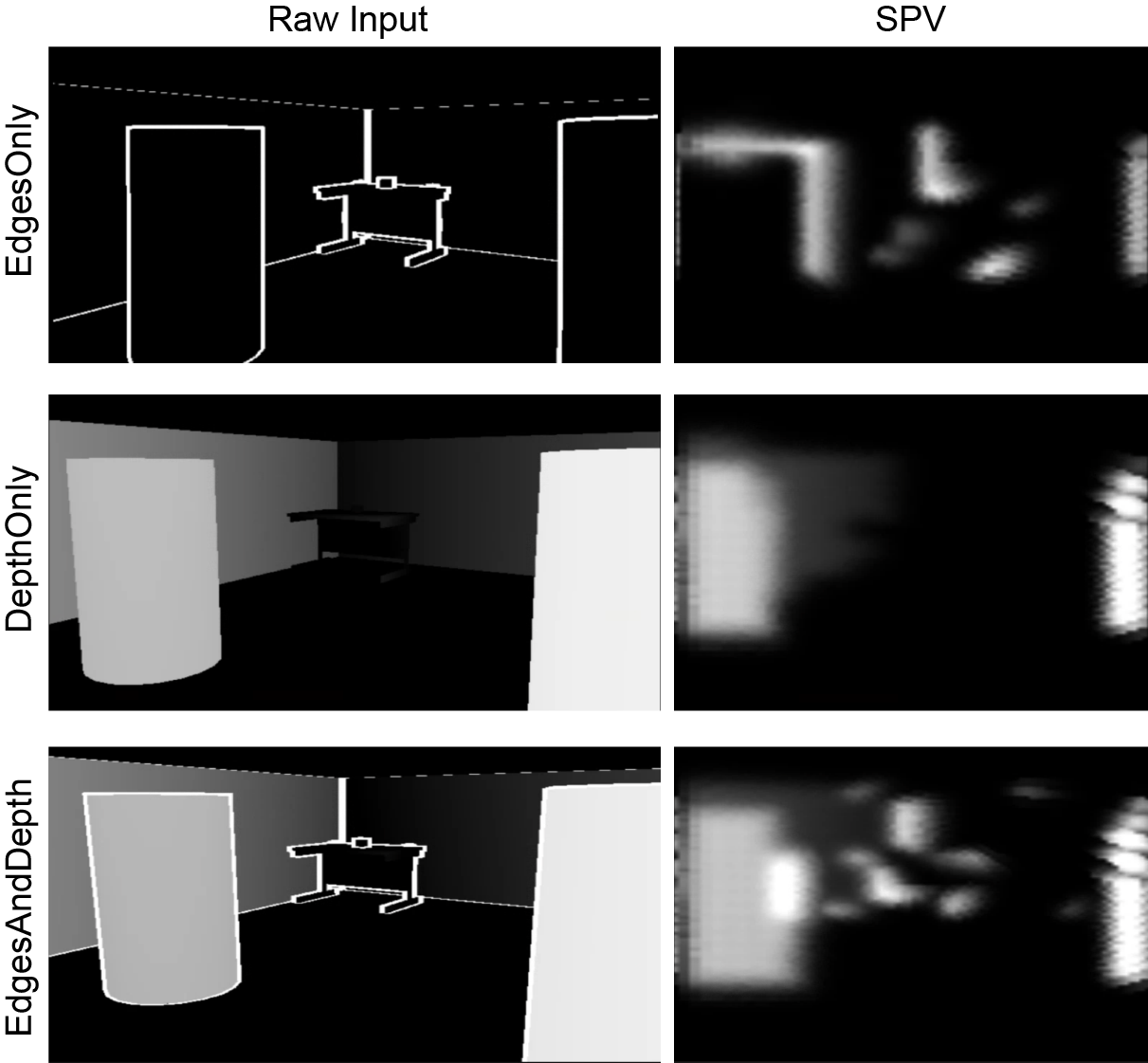}
    \Description{Example frames for the different modes, showing object outlines for EdgesOnly and shaded gray levels for DepthOnly. The scene includes two cylinders that are closeby and a table located further away in the corner of the room.}
    \caption{SPV modes used for scene simplification. \emph{Top:} In \emph{EdgesOnly} mode, only semantic and structural edges are visualized. \emph{Middle:} In \emph{DepthOnly} mode, per-pixel ground-truth depth is inverted and linearly translated to grayscale level. \emph{Bottom:} In \emph{EdgesAndDepth} both edges and depth are visualized. A fourth mode, \emph{EdgesOrDepth}, gave users the ability to toggle between \emph{EdgesOnly} and \emph{DepthOnly} modes by pressing a button on the VIVE controller.}
    \label{fig:modes}
\end{figure}

We considered four different scene simplification modes (Fig.~\ref{fig:modes}):
\begin{enumerate}[topsep=0em,itemsep=-1ex,partopsep=0em,parsep=1ex,leftmargin=6ex,label=\roman*.]
    \item \emph{EdgesOnly}: To simulate semantic and structural edge segmentation \cite{sanchez-garcia_semantic_2020}, we drew white outlines around the ground-truth object boundaries of the obstacles, tables, and objects in the room using the \texttt{QuickOutline} asset (Fig.~\ref{fig:modes}, \emph{top}).
    Rather than applying an edge extraction algorithm to the camera view (which may be too slow to run in real time),    \texttt{QuickOutline} coats each object with a white material, followed by a slightly smaller coat of a black material that makes it look as if only the edges of the object are showing (see Fig.~\ref{fig:modes}A for an example).
    In addition, we highlighted the edges of the room to give participants an indication of the room layout.
    \item \emph{DepthOnly}: To simulate depth to intensity substitution \cite{han_deep_2021}, we first extracted the ground-truth depth of each object relative to the main camera using \texttt{UnityObjectToClipPos}.
    Inverse depth was then linearly mapped to phosphene brightness, so that objects at zero depth appeared white, and objects at distances greater than the furthest wall (\texttt{farClipPlane}) appeared black (Fig.~\ref{fig:modes}, \emph{middle}). In other words, objects appeared brighter the closer they were to the observer.
    \item \emph{EdgesAndDepth}: This mode visualized both edge and depth information at the same time (Fig.~\ref{fig:modes}, \emph{bottom}).
    \item \emph{EdgesOrDepth}: In this mode, participants randomly started in either \emph{EdgesOnly} or \emph{DepthOnly} mode and were able to toggle between modes via press of a button on the VIVE controller.
\end{enumerate}

In our pilot study, we also included a low-resolution ``straight-through'' \ac{SPV} view of the scene (that did not include any scene simplification) as a baseline.
However, participants were unable to perform either task.
This is consistent with previous work showing that such a mode leads to chance performance in obstacle avoidance tasks (at best) \cite{kasowski_immersive_2022}.
To prevent participant frustration and to save time, we therefore did not include this mode in our final experiment.

\subsection{Procedure}
\label{sec:procedure}

The order in which the modes were presented and the order in which the rooms were presented was pseudo-randomized across participants.
To reduce cognitive load and context-switching, all rooms for a particular mode were presented in a block.
This allowed participants to get comfortable with one mode at a time.
Participants saw each combination of room and mode only once.

Before the start of each mode, participants were allowed to explore a tutorial room (not part of Rooms A--F) that contained a single obstacle to avoid and a medium-sized cube on a table without distractor objects.
Participants were free to navigate the room for as long as they wanted, until they felt comfortable with the \ac{SPV} mode.
When they indicated that they were ready to start the experiment, they had to grab the medium-sized cube off the table and confirm their selection with a button press.
They were then given $30$ seconds to walk back to the starting position, indicated by a red circle on the floor.
At this point the experiment started with a pseudo-randomly selected room.

A trial was concluded by selecting an object from one of the tables in the room.
We estimated from our pilot study that most participants take between thirty seconds and one minute per trial, but on rare occasions would get lost to the point where they could not complete a trial. We therefore introduced a time limit, which was set at four minutes per trial.
If participants exceeded this limit, the trial ended automatically, \ac{SPV} was turned off, and the participant was given $30$ seconds to walk back to the starting position, before the next trial started.

In \emph{EdgesOrDepth} mode, participants were allowed to toggle between \emph{EdgesOnly} and \emph{DepthOnly} modes as often as they wanted.
Each trial started in the mode that was last active during the last trial (or for Trial 1: whichever was last active in the tutorial room).
This way, if participants preferred one mode over the other, they could just stay on that mode without having to constantly switch back and forth between trials.

\begin{figure*}[t!]
    \centering
    \includegraphics[width=\textwidth]{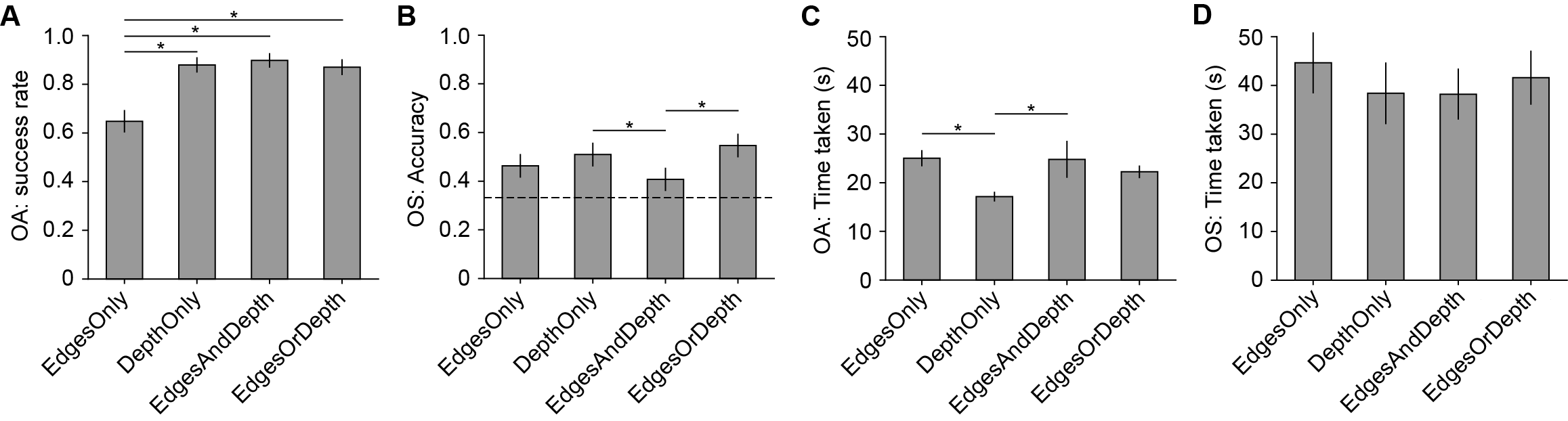}
    \Description{Barplots showing the success rates for obstacle avoidance and object selection. Numbers are explained in the main text. OA success rate was worst for EdgesOnly, with the three other modes performing similarly. OS accuracy was best for DepthOnly and EdgesOrDepth.
    Time taken was around 20s for OA and around 40s for OS. }
    \caption{Obstacle avoidance (OA) performance, measured by success rate (i.e., the number of trials with zero collisions, Panel A) and time taken (Panel C), and object selection (OS) performance, measured by accuracy (i.e., the fraction of trials where the correct object was selected, Panel B), and time taken (Panel D).
    The dashed line in Panel B indicates chance performance (\SI{33}{\percent}).
    Vertical bars are the standard error of the mean (SEM).
    Statistical significance was determined using paired $t$-tests, corrected for multiple testing using the Holm-Sidak method (*: $p<.05$).}
    \label{fig:results-overall}
\end{figure*}

\subsection{Data Collection and Analysis}

During the whole experiment, we recorded participant position (sampled every $0.5$ seconds), time and place of obstacle collisions, and object selection.
Upon collision with an obstacle, we applied a $3$ second timeout window so that prolonged collisions with the same object were only counted once.

We considered the number of collisions (the fewer the better) and time taken (the shorter the better) as the main metrics to quantify obstacle avoidance performance.
The obstacle avoidance part of the task ended as soon as participants entered the top third of the room.
From that time forward, participants were considered to be in the object selection part of the task, even if they accidentally crossed the line back into the obstacle avoidance portion of the room.

We considered the fraction of correctly selected objects (accuracy; the higher the better) and time taken (the shorter the better) as the main metrics to quantify object selection performance.
A trial ended as soon as participants selected an object.


To determine whether performance was significantly different across modes ($C(\mathit{Mode})$) and rooms ($C(\mathit{Room})$), we ran a regression analysis for each metric mentioned above using the Ordinary Least Squares (OLS) model in the \texttt{statsmodel} Python module.
We also factored in participant's gender ($C(\mathit{Gender})$) and VR experience ($C(\mathit{VRExperience})$).
To test for learning effects, we labeled trials depending on whether they were completed in the first or second half of each session ($C(\mathit{FirstHalf})$).
The full model was thus $\mathit{Metric} \sim C(\mathit{Mode}) + C(\mathit{Room}) + C(\mathit{Gender}) + C(\mathit{VRExperience}) + C(\mathit{FirstHalf})$.
$p$-values were calculated using paired $t$-tests of the OLS model, which corrects for multiple testing using the Holm-Sidak method.

\section{Results}

All participants were able to complete the experiment using all four tested modes, taking on average $25 \pm 15$ minutes to complete the 24 different trials.
Trial completion time varied widely across rooms and modes, with a median completion time of 16 seconds for the obstacle course, 12 seconds for the object selection portion, and 31 seconds overall.
90\% of trials were completed in less than 193 seconds, with the obstacle course cleared in under 35 seconds and the object selected in under 155 seconds.
The time limit of 4 minutes was reached on only two trials, which were subsequently removed from further analysis.

\subsection{Obstacle Avoidance}

The number of obstacle avoidance (OA) trials completed without any collisions (``success rate'') was significantly affected by the scene simplification mode ($F(3, 432) = 32.2, p<.001$) and room layout ($F(5, 432) = 2.49, p<.05$), summarized in Fig.~\ref{fig:results-overall}A.
Participants performed the OA portion of the task most successfully using the \emph{EdgesAndDepth} mode, completing on average \SI{89.8}{\percent} of trials without colliding.
However, this was not significantly better than with the \emph{DepthOnly} (\SI{87.9}{\percent}) and \emph{EdgesOrDepth} modes (\SI{87.0}{\percent}; $t$-test $p>0.05$, corrected for multiple comparisons with the Holm-Sidak method).
On the other hand, participants performed worst with the \emph{EdgesOnly} mode, as evidenced by a significantly lower success rate (\SI{64.8}{\percent}) and among the longest times taken (\SI{25}{\second}, Fig.~\ref{fig:results-overall}C).
Participants were significantly faster using \emph{DepthOnly} (\SI{17}{\second}) than they were with \emph{EdgesOrDepth} (\SI{22}{\second}) and \emph{EdgesAndDepth} (\SI{25}{\second}; Fig.~\ref{fig:results-overall}C).

In addition, we found a strong learning effect ($F(1, 432)=84.3$, $p<.001$), which indicated an increased success rate of 28\% from the beginning of a session to the end.
Weaker but significant effects were also found for gender ($F(1, 432) = 18.6, p<.001$) and VR experience ($F(3, 432) = 3.38, p<.05$).

\begin{figure*}[tb]
    \centering
    \includegraphics[width=\linewidth]{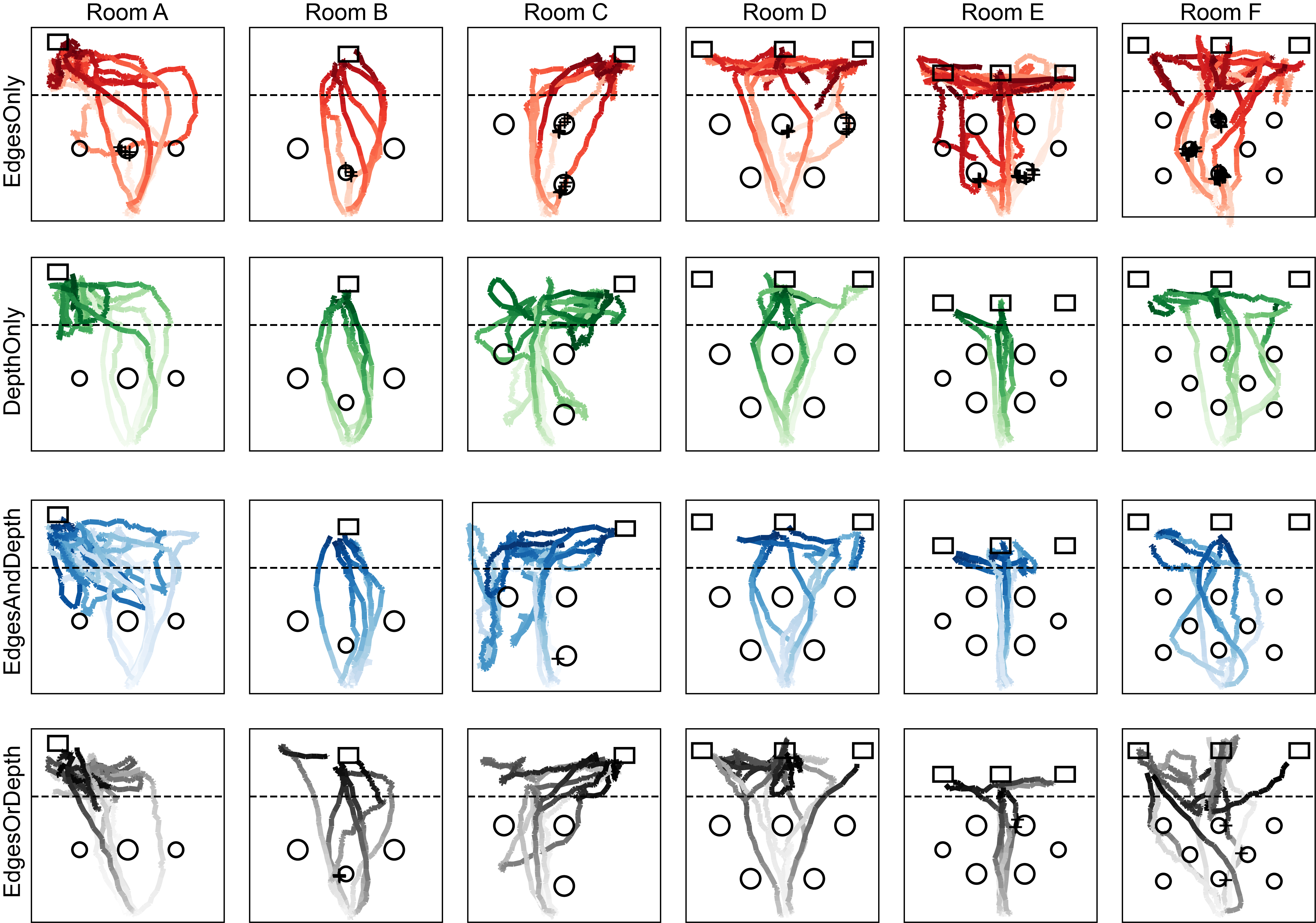}
    \Description{Example paths taken for all rooms and modes. EdgesOnly led to most collisions. Room F led to most collisions. }
    \caption{Birds-eye view of paths taken in the different rooms (columns) using the different scene simplification strategies (rows).
    For the sake of clarity, only the paths from every fourth subject are shown.
    Color of paths gets more saturated as time moves on.
    Collisions are indicated with a black \texttt{+} sign.
    Circles indicate the location of the obstacles, and rectangles are the table.
    The task switched from obstacle avoidance to object selection as soon as the participant crossed the (to them invisible) horizontal dashed line.}
    \label{fig:results-paths}
\end{figure*}

\subsection{Object Selection}

The number of object selection (OS) trials with correct object selections was significantly affected by the scene simplification mode ($F(3, 432) = 6.85, p<.001$) and room layout ($F(5, 432) = 10.2, p<.001$), summarized in Fig.~\ref{fig:results-overall}B.
Participants performed the OS portion of the task most successfully using the \emph{EdgesOrDepth} mode, on average locating and selecting the correct object in \SI{54.6}{\percent} of trials (Fig.~\ref{fig:results-overall}B).
However, this was not significantly better than with the \emph{EdgesOnly} (\SI{46.3}{\percent}) and \emph{DepthOnly} modes (\SI{50.9}{\percent}).
Performance with the \emph{EdgesAndDepth} mode (\SI{40.7}{\percent}) was only slightly better than chance (\SI{33}{\percent}).

OS performance was also affected by gender ($F(1, 432) = 16.2, p<.001$) and VR experience ($F(3, 432) = 5.01, p<.001$).
However, we did not find any learning effects here ($F(1, 432) = .48, p=.488$).

All participants took approximately \SI{40}{\second} to walk towards the correct table and select the correct object, no matter which scene simplification mode was used (Fig.~\ref{fig:results-overall}D).
Since the location of the correct object was pseudo-randomized on each trial, it is possible that most of the time for the OS portion of the task was spent simply walking back and forth between the three tables.

\subsection{Participant Paths}

The paths that participants took are summarized in Fig.~\ref{fig:results-paths}, showing a birds-eye view of the six different rooms.
Here, each line represents the path of a participant, with colors getting more saturated as time progresses, and collisions indicated by black crosses.

The poor OA performance of the \emph{EdgesOnly} mode first reported in Fig.~\ref{fig:results-overall} can also be appreciated here, with participants causing more collisions compared to the other modes across all rooms.
This is especially evident in Room F: Even though participants attempted to avoid the obstacle maze by moving in a slalom fashion, there are only two obstacles that never led to collisions.

It is interesting to note that participants seemed to explore Room A the most, with some participants accidentally walking away from the table in the top-left corner of the room and returning to the OA portion of the task.
On the other hand, most participants walked mostly straight ahead in Room B (except in \emph{EdgesOrDepth} mode) and Room E, thus avoiding most obstacles, although most participants still collided with the obstacle that was closest to the starting position.
Room F was by far the most challenging, leading to a large number of collisions in all four modes.

\subsection{Relative Importance of Depth vs. Edges}
One potential indicator of the relative importance of depth and edge cues was \emph{EdgesOrDepth} mode: here participants had the ability to switch between \emph{EdgesOnly} and \emph{DepthOnly} information at the push of a button as often as they saw fit.

\begin{figure}[!t]
    \centering
    \includegraphics[width=\linewidth]{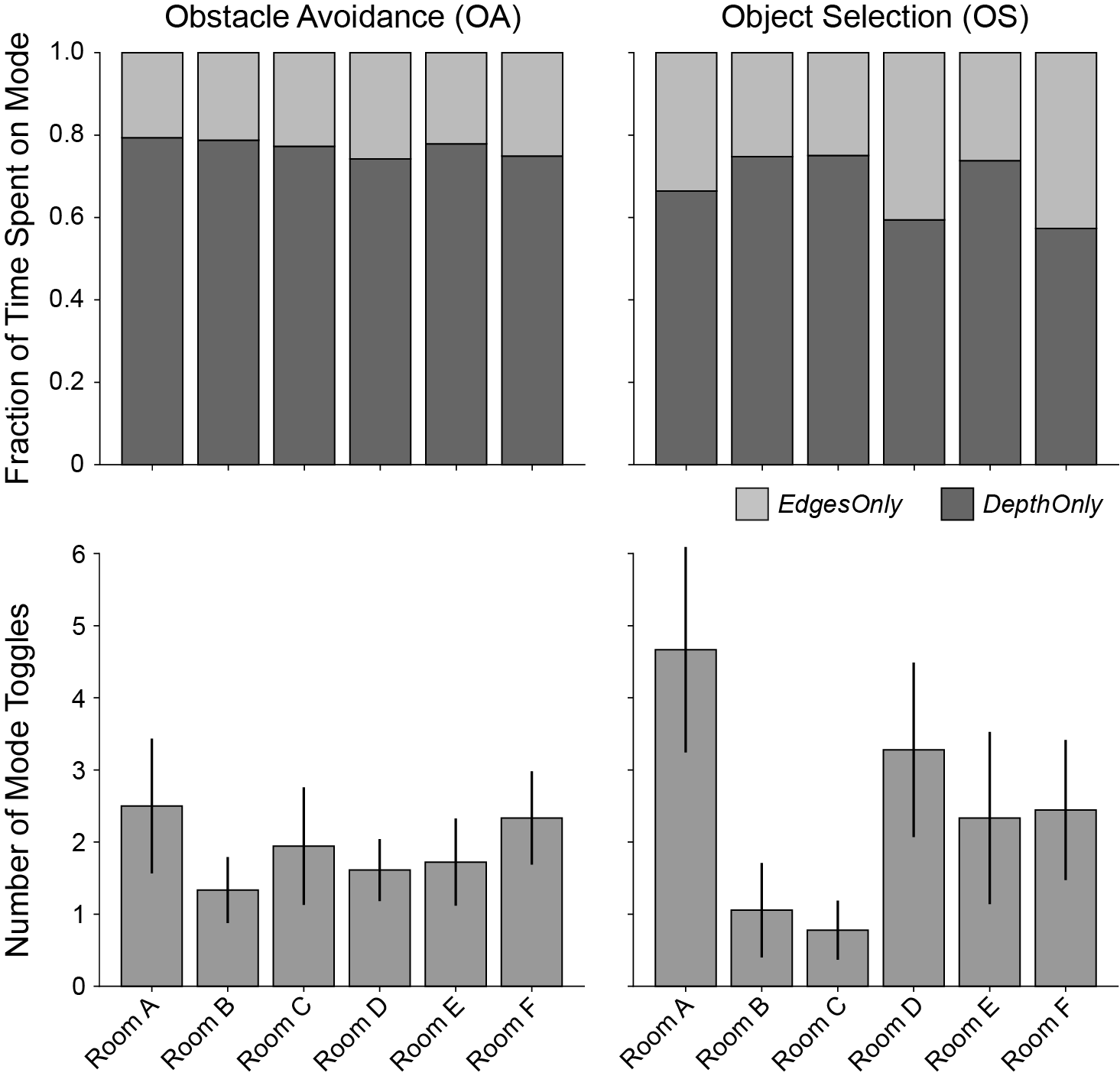}
    \Description{Barplots describing user behavior during EdgesOrDepth mode. Top: barplots indicating that most time was spent on DepthOnly mode, in all rooms, during both obstacle avoidance and object selection. Bottom: Barplots indicating how often users toggled between EdgesOnly and DepthOnly mode across the six rooms. In obstacle avoidance, users switched between 1 and 2 times in all rooms. In object selection, people switched most in Room A.}
    \caption{\emph{EdgesOrDepth}: Fraction of time spent on each mode (\emph{top}) and number of mode toggles (\emph{bottom}) for each room, for the obstacle avoidance portion of the task (\emph{left}) and the object selection portion of the task (\emph{right}).}
    \label{fig:results-switching}
\end{figure}

We recorded both the number of times that participants toggled modes as well as the amount of time spent in each  mode. The results are shown in Fig.~\ref{fig:results-switching}.
To our surprise, participants spent most of their time in \emph{DepthOnly} mode (Fig.~\ref{fig:results-switching}, \emph{top}), both during the OA as well as the OS portion of the task.

In addition, most participants did not often switch between modes ((Fig.~\ref{fig:results-switching}, \emph{bottom}).
The experiment was set up so that each trial started on the mode that was last active (either during the preceding trial or in the tutorial room), so that participants who preferred one mode over the other would not have to constantly switch back and forth.
During the OA portion of the task, participants tended to toggle modes either never (perhaps because the trial started in their preferred mode), once (perhaps because they trial started in their nonpreferred mode), or twice (perhaps to briefly view the other mode, but then promptly switch back to the preferred one).

The only exception was Room A, where participants tended to switch modes much more often. Since the orders of the rooms was pseudo-randomized, it is not entirely clear why Room A would prompt participants to switch modes more often than other rooms.

\subsection{User Preferences}
Upon completion of the experiment, we asked participants which mode they preferred for the OA portion of the task, the OS portion of the task, and overall (Fig.~\ref{fig:results-preference}).
Consistent with Fig.~\ref{fig:results-overall}A, participants preferred  \emph{DepthOnly} and \emph{EdgesOrDepth} to avoid obstacles; $\chi^2(1, N=18) = 8.22, p<.05$.
However, \emph{EdgesAndDepth} also resulted in strong OA performance, yet was only preferred by two participants.
Most participants preferred the ability to switch between depth and edge cues by means of the \emph{EdgesOrDepth} mode.

When it came to selecting objects, user preferences were mixed (Fig.~\ref{fig:results-preference}, \emph{center}) and not significantly difference from a uniform random distribution: $\chi^2(1, N=18)=.667, p=.88$.
This is consistent with the somewhat lackluster performance of participants in the object selection portion of the task (Fig.~\ref{fig:results-overall}B), suggesting that none of the tested modes may be ideal for object recognition.

Lastly, overall preference mirrored the OA result, with most participants preferring \emph{DepthOnly} and \emph{EdgesOrDepth} mode (Fig.~\ref{fig:results-preference}, \emph{right}).
However, this result was not significantly different from a random uniform distribution, although it approached significance ($\chi^2(1, N=18)=7.33, p=.06$).
Notably, none of the subjects preferred \emph{EdgesOnly} mode overall, and only four subjects preferred \emph{EdgesAndDepth} mode.
The majority of participants was split between being comfortable using \emph{DepthOnly} mode and preferring the flexibility to switch between modes using \emph{EdgesOrDepth} mode (even though most time was spent looking at depth cues).

Overall these results further corroborate the relative importance of depth cues for both obstacle avoidance and object identification.

\section{Discussion}

\subsection{Relative Importance of Depth Cues and Semantic Edges for Bionic Vision}

Here we asked sighted participants to act as virtual bionic eye recipients by navigating virtual indoor environments using simulated prosthetic vision.
To the best of our knowledge, this is the first study that directly compares the relative importance of depth cues and semantic edges for bionic vision. 

\begin{figure*}[!bt]
    \centering
    \includegraphics[width=\linewidth]{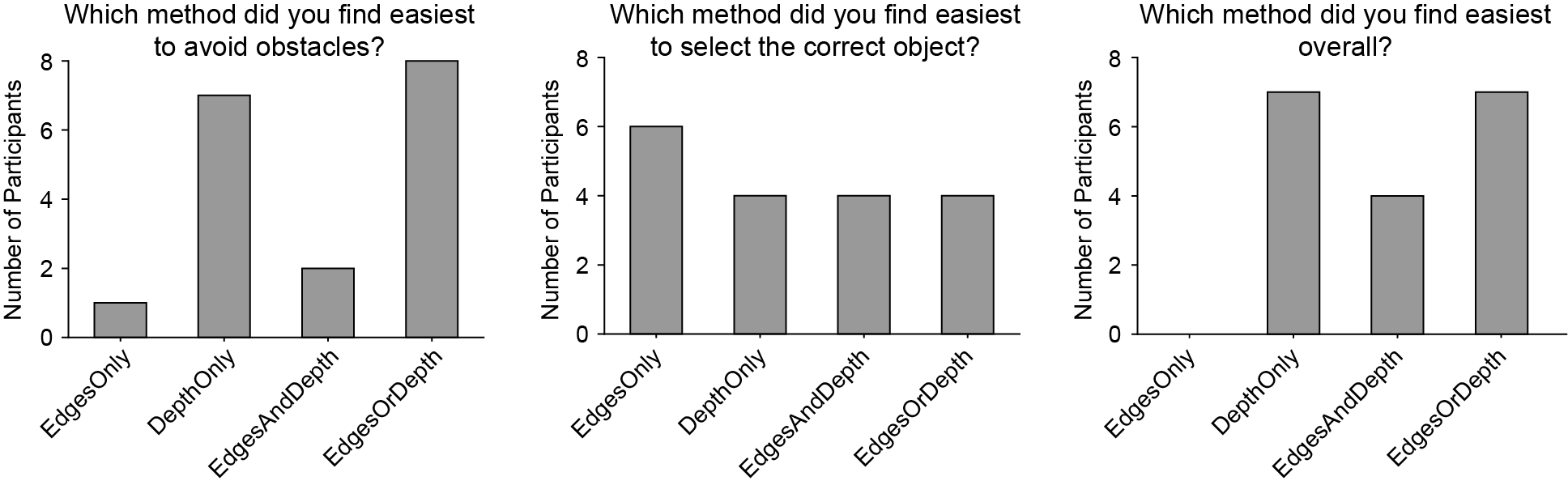}
    \Description{Barplots of user preferences for the four different modes in three panels.
    In the leftmost panel, users indicate their preferred mode to avoid obstacles: 8 users preferred EdgesOrDepth, 7 users preferred DepthOnly, 2 users preferred EdgesAndDepth, and 1 user preferred EdgesOnly.
    In the middle panel, users indicate their preferred mode to select objects: 6 users preferred EdgesOnly mode, and the other modes received 4 votes each.
    In the rightmost panel, users indicate their preferred mode overall: 7 users preferred DepthOnly, 7 users preferred EdgesOrDepth, 4 users preferred EdgesAndDepth, and no users preferred EdgesOnly.}
    \caption{Reported user preferences for the different SPV modes.}
    \label{fig:results-preference}
\end{figure*}

Boxplots of user response times in milliseconds for the six experimental conditions. Mean values are provided in Table 1, column 4. Interquartile ranges are typically 2-3 seconds, with the Gesture+Voice condition having a broader range of 5 seconds and the smallest being the Voice condition at 2 seconds. Voice and Face have the lowest mean values and low standard deviations. Gesture+Voice has the highest. Face+Voice has few outliers.

We found that participants were significantly better at avoiding obstacles using depth-based methods as opposed to relying on edge information alone (Fig.~\ref{fig:results-overall}A).
These results are consistent with previous studies noting the importance of highlighting nearby obstacles for orientation \& mobility tasks \cite{mccarthy_mobility_2014,lieby_substituting_2011}.
However, the exact way in which depth cues were presented seemed to carry little importance, as evidenced by similar performance of the \emph{DepthOnly}, \emph{EdgesAndDepth}, and \emph{EdgesOrDepth} modes.
Nevertheless, it is interesting that participants completed the obstacle avoidance portion of the task the fastest using \emph{DepthOnly} mode (Fig.~\ref{fig:results-overall}C), even though \emph{EdgesAndDepth} provided the same amount of depth cues at all times.

Theoretically, participants should have performed (asymptotically) the same with the \emph{EdgesOrDepth} mode, as it provides at least as much information as \emph{DepthOnly} mode.
Its purpose was to provide all the functionality of the individual modes (i.e., \emph{DepthOnly} and \emph{EdgesOnly}) while giving the user the power and flexibility to choose whichever mode they deemed best for a particular situation.
Instead, this freedom may have added unnecessary overhead (in terms of time taken to complete the task) and opened up the possibility that users may perform the task in a suboptimal mode.
This is an important \ac{HCI} insight to consider for future designs of bionic eyes that rely on computer vision for scene simplification \cite{beyeler_towards_2022}.

\subsection{Depth and Semantic Edges May Not Be Sufficient for Object Recognition}

Depth cues were also important for object selection (Fig.~\ref{fig:results-overall}B), with participants performing best using the \emph{DepthOnly} and \emph{EdgesOrDepth} modes.
However, performance was more lackluster as compared to the obstacle avoidance portion of the task.
Participants performed near chance levels using \emph{EdgesAndDepth}, and were only slightly better with the other three modes.
Only \emph{DepthOnly} and \emph{EdgesOrDepth} were significantly better than \emph{EdgesAndDepth}, and time taken to complete this subtask was similar across modes.

Overall these results suggest that presenting both edge and depth cues at once may hinder participants' ability to segregate important objects from the background.
Yet none of the tested modes seem to provide sufficient information for participants to excel at this portion of the task.

\subsection{Users Prefer Depth Information Over Edges}

The relative importance of depth information was further corroborated by Fig.~\ref{fig:results-switching}, which indicates that when participants were given a choice between edges and depth, they largely preferred to navigate the room by relying on depth information alone.

In obstacle avoidance and overall, most participants preferred \emph{DepthOnly} and \emph{EdgesOrDepth} over the other two modes (Fig.~\ref{fig:results-preference}), but opinions were split:
Approximately half of participants preferred the flexibility of switching between edges and depth, whereas the other half was content with \emph{DepthOnly} mode.
The lackluster object selection performance is also evidenced by the user preference chart, which was not significantly different from random.

Overall these results suggest that while depth information seems to be the most beneficial for obstacle avoidance and object selection, there may not be one scene simplification made that is preferred by the majority of bionic eye recipients.

\subsection{Limitations and Future Work}

Although this study effectively uses immersive \ac{VR} to address previously unanswered questions about \ac{SPV}, there are several limitations that should be addressed in future work as outlined below.

First, it is important to point out that the study was performed on sighted undergraduate students readily available at UCSB.
Their age, navigational affordances, and experience with low vision may therefore be drastically different from real bionic eye recipients, who not only tend to be older (and prolific cane users), but also receive extensive vision rehabilitation training.

Second, we acknowledge that the navigating with simulated prosthetic vision was a challenging task even for experienced users of \ac{VR}.
As a result, there were isolated cases of participants becoming reckless and charging straight ahead through the obstacle course or simply selecting the first object they encountered.
On such occasion, behavior in the virtual environment would thus be noticeably different from a real-world obstacle course, where collisions have physical consequences.
A next step in this line of research would thus be to provide the different scene simplification modes in a real-world environment.
Depth information could be collected with an RGB-D camera and the different \ac{SPV} modes could be rendered on an augmented-reality \ac{HMD}.
While an immersive virtual environment has clear advantages (e.g.,  safety, access to ground-truth depth information, perfect control over the layout and illumination of the visual environment), behavioral performance of virtual patients in a real-world obstacle course might be closer to the application that this work would eventually feed into.
The ultimate test, of course, would be to demonstrate that these scene simplification modes can support everyday tasks of real bionic eye users.

Interestingly, we found vast individual differences in task performance, which were not unlike those reported in the literature \cite{he_improved_2020}.
Subjects who did well with one mode tended to do well across all modes (data not shown), suggesting that some people were inherently better at adapting to prosthetic vision than others.
Future work should therefore zero in on the possible causes of these individual differences and compare them to real bionic eye users.
Studying these differences could identify training protocols that can enhance the ability of all device users.

\section{Conclusion}

In sum, this is the first study to compare the relative importance of depth cues and semantic edges for bionic vision and constitutes an essential first towards a bionic eye that uses computer vision to improve a user's scene understanding.

\begin{acks}
This work was supported by the National Institutes of Health (NIH R00 EY-029329 to MB).
\end{acks}

\bibliographystyle{ACM-Reference-Format}
\bibliography{2022-VRST-Indoor-Navigation}






\end{document}